\documentclass[aps,twocolumn,superscriptaddress,amsfont,amssymb,amsmath,showpacs,balancelastpage,english]{revtex4-1} 
\usepackage{amsmath}
\usepackage{amssymb}
\usepackage{babel}
\usepackage{graphicx}
\usepackage{color}
\usepackage{natbib}
\usepackage{longtable}



\begin{document}

\title{Measurement of Hubble constant with stellar-mass binary black holes}

\author{Atsushi Nishizawa}
\email{anishi@kmi.nagoya-u.ac.jp}
\affiliation{Kobayashi-Maskawa Institute for the Origin of Particles and the Universe, Nagoya University, Nagoya 464-8602, Japan}
\affiliation{Department of Physics and Astronomy, The University of 
Mississippi, University, MS 38677, USA}

\begin{abstract}%%%%%%%%%%%%%%%%%%%%%%%%%%%%%%%%%%%%%%%
The direct detections of gravitational waves (GW) from merging binary black holes (BBH) by aLIGO have brought us a new opportunity to utilize BBH for a measurement of the Hubble constant. In this paper, we point out that there exists a small number of BBH that gives significantly small sky localization volume so that a host galaxy is uniquely identified. Then a redshift of a BBH is obtained from a spectroscopic follow-up observation of the host galaxy. Using these redshift-identified BBH, we show that the Hubble constant is measured at a level of precision better than $1\%$ with advanced detectors like aLIGO at design sensitivity. Since a GW observation is completely independent of other astrophysical means, this qualitatively new probe will help resolve a well-known value discrepancy problem on the Hubble constant from cosmological measurements and local measurements.
\end{abstract}

\date{\today}

\maketitle

%%%%%%%%%%%%%%%%%%%%%%%%%%%%%%%%%%%%%%%%%%%%

\section{Introduction}
% Hubble constant discrepancy problem
It is well known that there is a discrepancy between the values of the Hubble constant determined from cosmological measurements such as cosmic microwave background (CMB) \cite{Planck2015cosmology} and baryon acoustic oscillation (BAO) \cite{Beutler:2011MNRAS} and local measurements using Cepheid variables \cite{Riess:2016ApJ} (for a review, see \cite{Jackson:2015LRR}). This discrepancy could be caused by a systematic error in the measurements or by dark radiation, which is unknown additional radiation and increases the number of relativistic species in the early Universe \cite{Planck2015cosmology}. In any case, pinning down the Hubble constant is crucial for understanding the standard model of cosmology, and requires another independent measurement qualitatively different from ones above. 

% GW observation
A gravitational-wave (GW) observation provides a new opportunity to measure the Hubble constant. The direct detections of GW from merging binary black holes (BBH) during the observation runs of aLIGO \cite{GW150914:detection,GW151226:detection,GW170104:detection} have demonstrated that the advanced detectors have sufficient sensitivity enough to detect GW out to the distant Universe. The three events detected so far, plus one candidate, also suggest that BBH mergers are common in the Universe, as already predicted before the detections in \cite{Belczynski:2010ApJL,Belczynski:2015ApJ}. These facts allow us to use BBH as a cosmological probe. The observation of GW from a compact binary gives luminosity distance to the source directly without any help of a distance ladder. Given source redshift information, the compact binaries can be utilized for measuring the cosmic expansion \cite{Schutz:1986Nature} and in this cosmological context they are called {\it the standard sirens} \cite{Holz:2005ApJ}. However, availability of the standard siren depends on whether a source redshift is available or not, because GW observation alone is not sensitive to the source redshift. If compact binaries are double neutron star (NS) binaries or NS-BH binaries, it is often assumed that the source redshift is obtained from an electromagnetic counterpart that occurs coincidentally with the GW event \cite{Sathyaprakash:2009xt,Nissanke:2013fka,Tamanini:2016JCAP}, though the coincidence rate is still largely uncertain \cite{Yonetoku:2014fua}. The other ways to obtain redshift information are assuming equation of state of a NS \cite{Messenger:2011gi} or a narrow mass distribution of NS \cite{Taylor:2011fs}. On the other hand, we cannot expect an electromagnetic counterpart for stellar-mass BBH nor use nongravitational properties of NS to obtain redshift information. 

% methods applicable to BBH
There have been two methods for BBH observed by ground-based detectors that do not resort to identifying electromagnetic counterparts. First one is a statistical method assuming a source redshift distribution based on galaxy catalogs \cite{MacLeod:2008PRD,Petiteau:2011ApJ,DelPozzo:2012PRD}. Each GW event has typically large sky error volume that contains many candidates of source host galaxies. By combining a large number of sources, a set of cosmological parameters consistent with all GW events is chosen. This method, however, can only be applied to GW sources at low redshifts, $z \lesssim 0.1$, because no galaxy catalog is complete in realistic observations at higher redshifts unless an intentional follow-up galaxy survey dedicated for GW events is performed in the future \cite{Gehrels:2015ApJ}. Second method is to utilize anisotropies of GW events on the sky \cite{Namikawa:2016PRL,Namikawa:2016PRD,Oguri:2016PRD}. The spatial distribution of BBH is anisotropic if they trace galaxy clustering induced by the large-scale structure of the Universe. The anisotropic signal contains rich cosmological information helpful to constrain the cosmic expansion history and structure formation without redshift information. 

In this paper, we focus on BBH observed by aLIGO-like detectors and show that the first method above for BBH is utilized to measure a local rate of the cosmic expansion, that is, the Hubble constant, at an unprecedented precision, $< 1\%$. It is remarkable that there exists a small number of BBH that gives significantly small sky localization volume containing a unique host galaxy. This new opportunity can be revealed only by statistically studying the parameter estimation errors of stellar-mass BBH from large samples of 50000, taking into account an astrophysical mass distribution, a redshift distribution, the realistic merger rate of BBH, and the up-to-date phenomenological waveform of GW. Then once a unique host galaxy is identified, the redshift of a BBH is obtained from a spectroscopic follow-up observation of the host galaxy at later time. In this sense, our result is conservative in that any galaxy catalog is not assumed {\it a priori}. The main conclusion of this paper is that the Hubble constant can be measured at less than 1\% level even with second-generation detectors like aLIGO, with better precision than other astrophysical means, independent of astrophysical systematics in other astrophysical sources.

% contents of this paper
This paper is organized as follows. In Sec.~\ref{sec2}, we begin with generating a source catalog with Monte Carlo method, taking into account an astrophysical situation as realistic as possible: mass distribution, redshift distribution, and networks of realistic GW detectors. In Sec.~\ref{sec3}, we estimate model parameters of BBH and compute error volume of sky localization for each source. Then we count the expected number of host galaxy candidates in the volume. Based on the number of GW sources with a unique host galaxy, we estimate the measurement precision of Hubble constant in Sec.~\ref{sec4}. Finally, Sec.~\ref{sec5} and \ref{sec6} are devoted to discussion and summary. Throughout the paper, we adopt units $c=G=1$.

%%%%%%%%%%%%%%%%%%%%%%%%%%%%%%%%

\begin{figure*}[t]
\begin{center}
\raisebox{0mm}{\includegraphics[width=5.5cm]{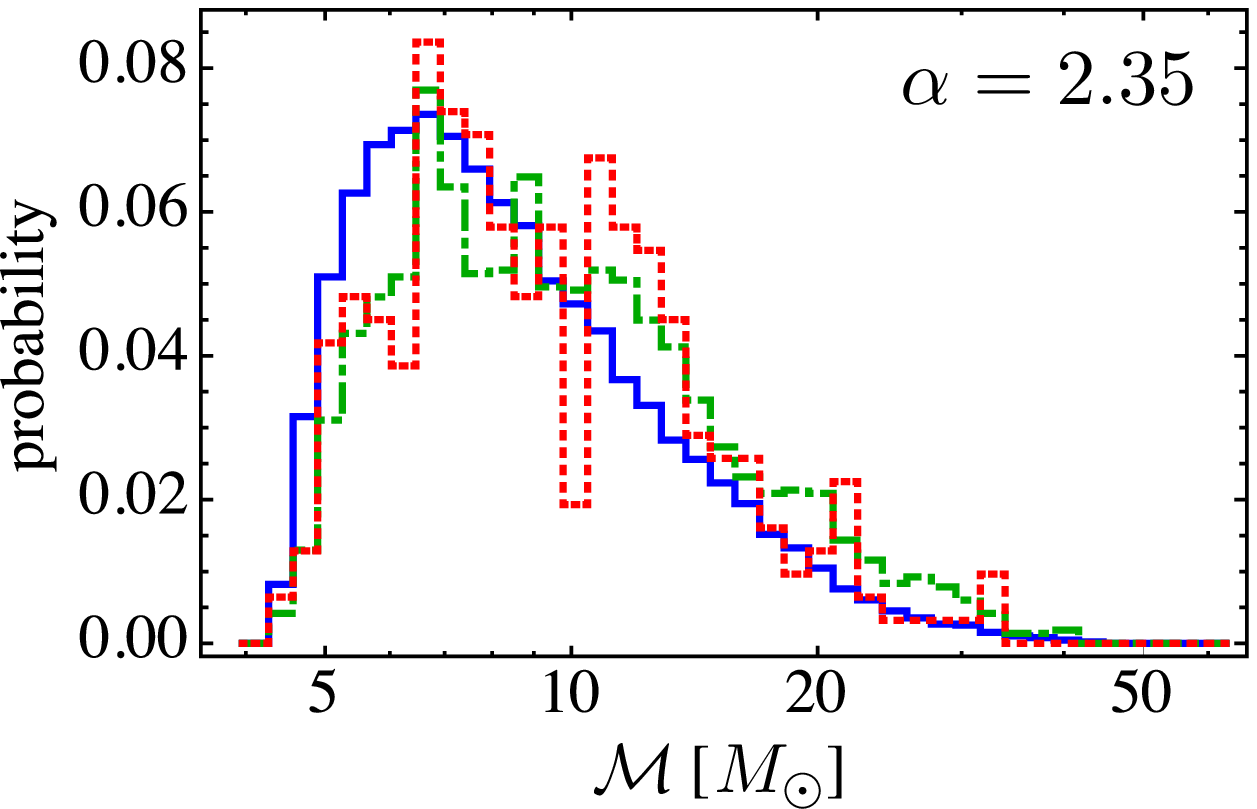}}
\hspace{2mm}
\raisebox{2.8mm}{\includegraphics[width=5.5cm]{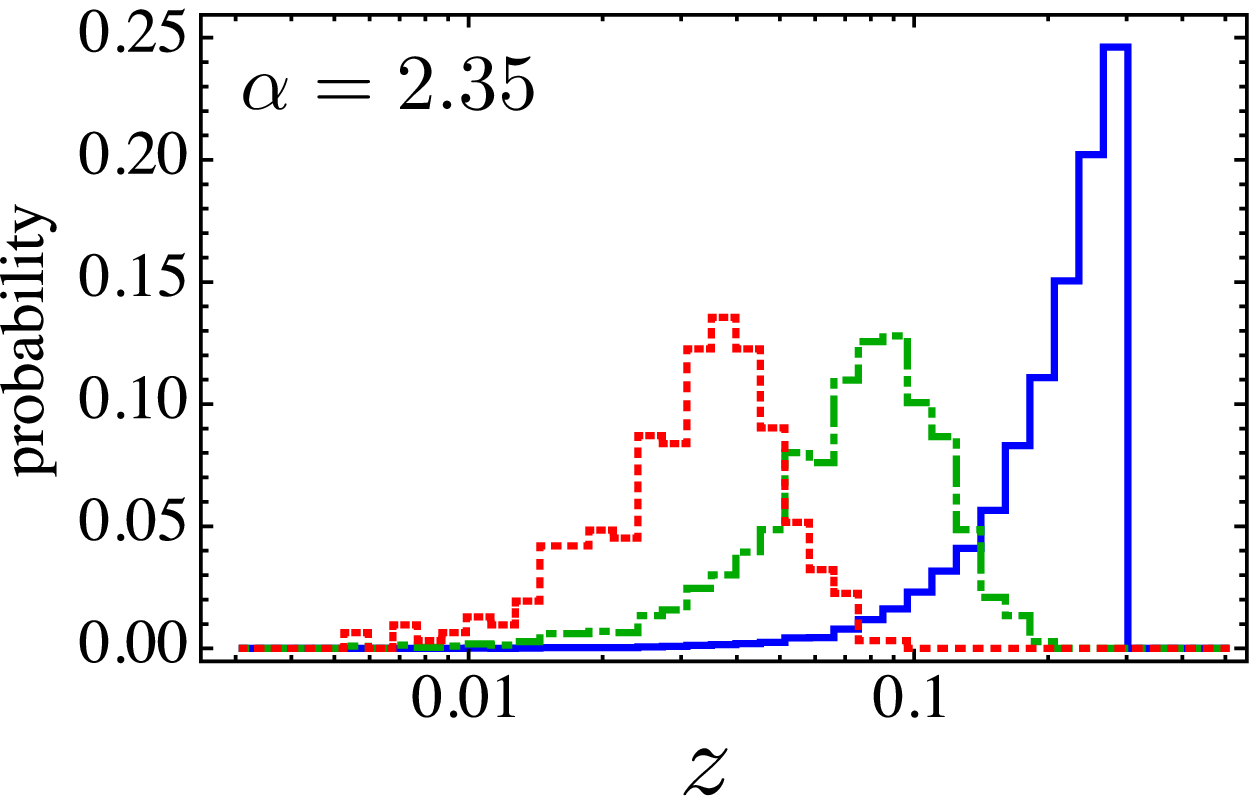}}
\hspace{2mm}
\raisebox{4.5mm}{\includegraphics[width=5.5cm]{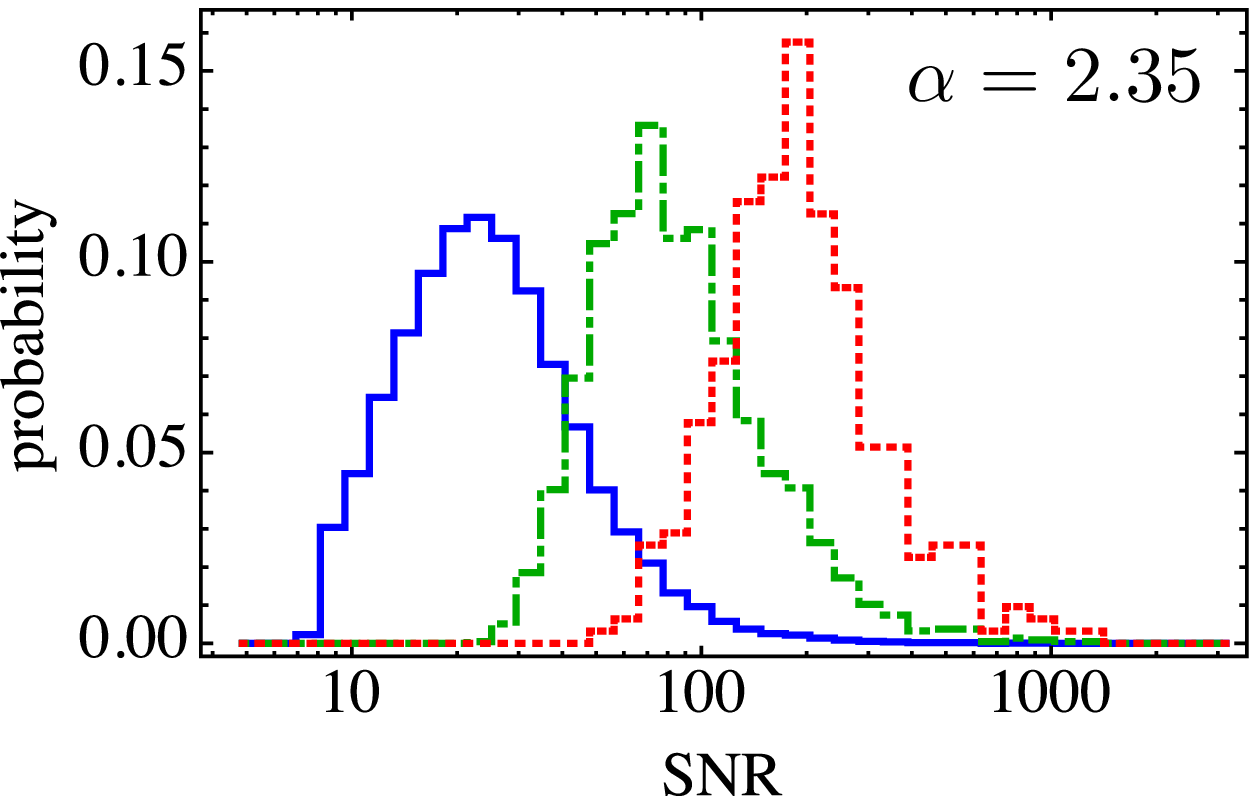}}
\raisebox{0mm}{\includegraphics[width=5.5cm]{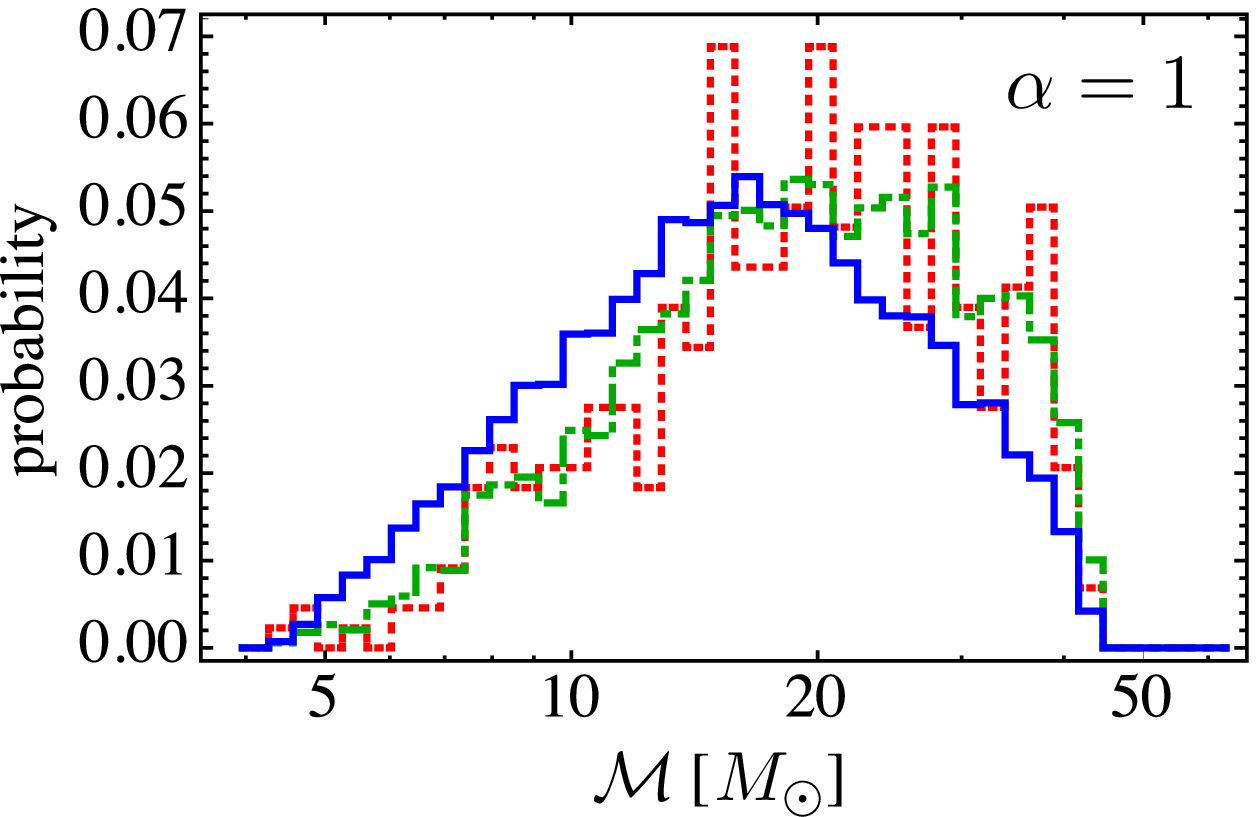}}
\hspace{2mm}
\raisebox{2.8mm}{\includegraphics[width=5.5cm]{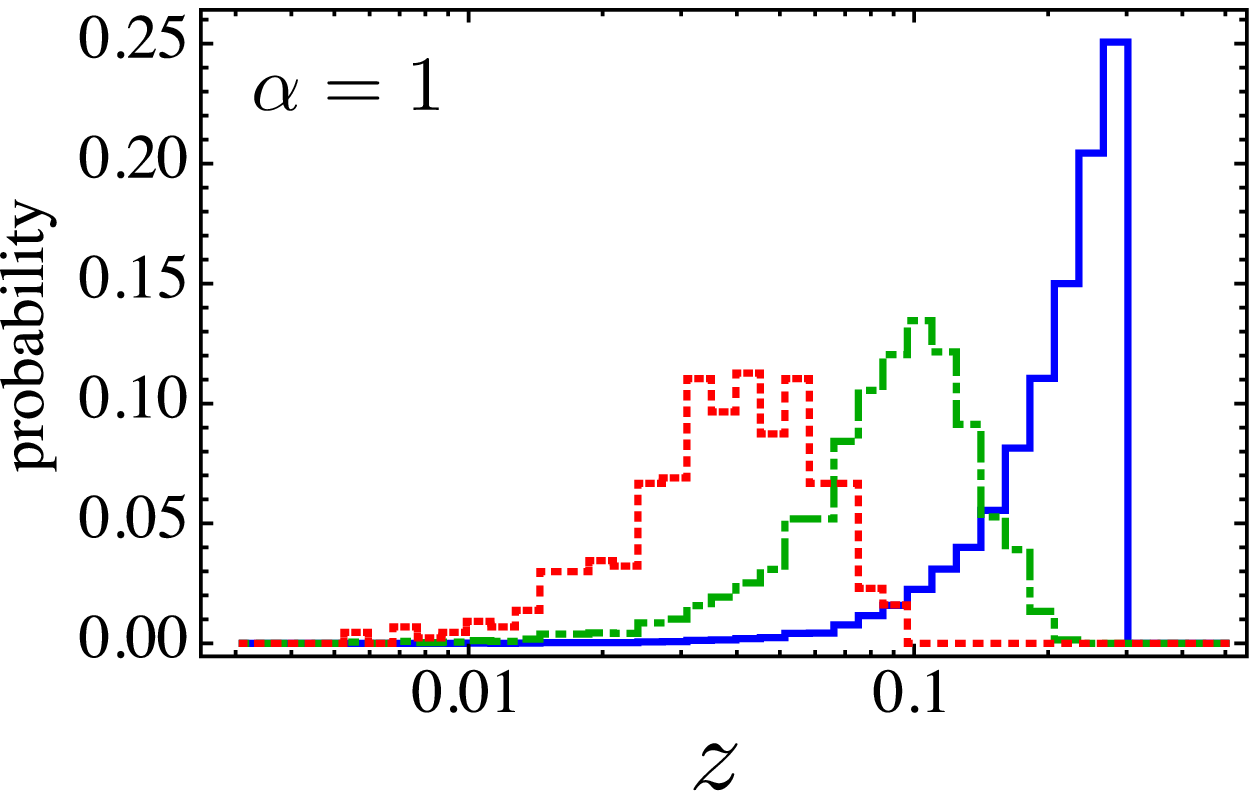}}
\hspace{2mm}
\raisebox{4.5mm}{\includegraphics[width=5.5cm]{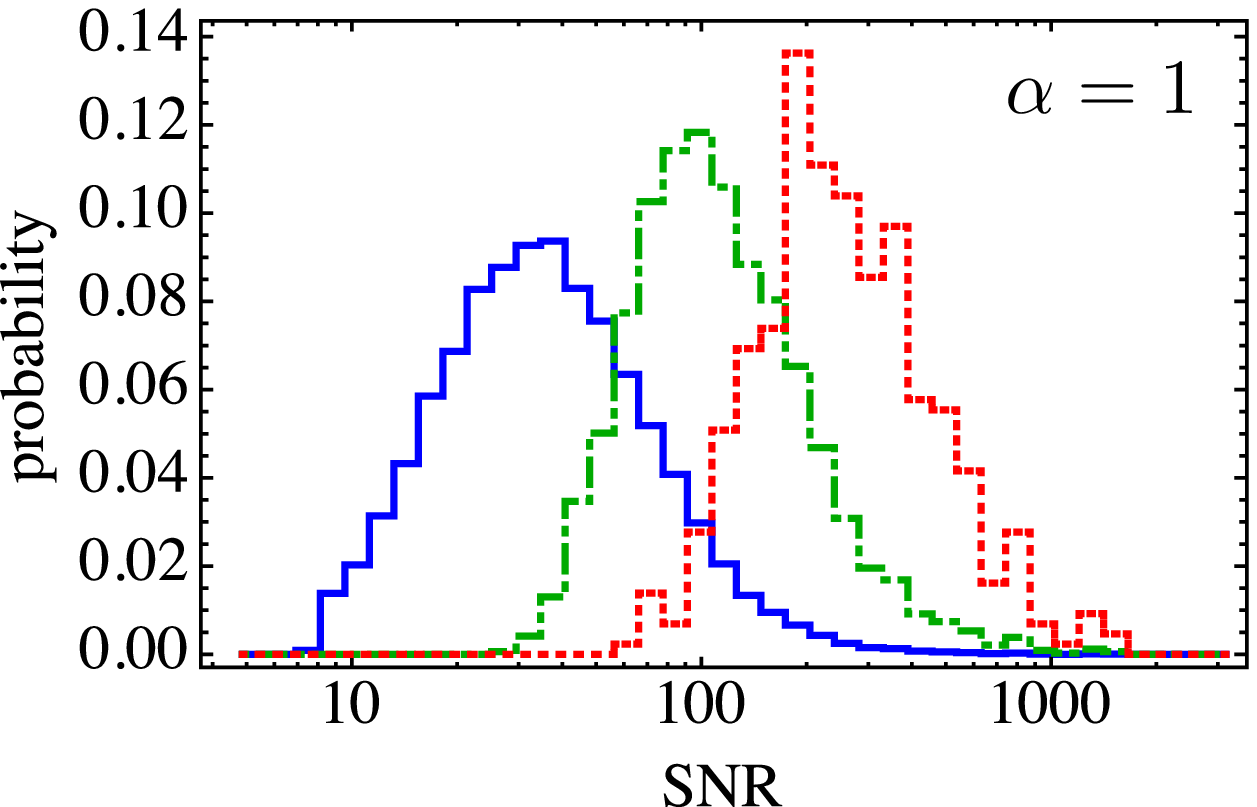}}
\caption{Probability distributions of source parameters in a source catalog with the mass distributions ($\alpha=2.35$ and $\alpha=1$), observed by the HLV detector network. The horizontal axes are from the left, the chirp mass, redshift, and SNR. The colors are all binaries (blue, solid), binaries with $N_{\rm host}<100$ (green, dot-dashed), and binaries with $N_{\rm host}<1$ (red, dotted).}
\label{fig1}
\end{center}
\end{figure*} 

\section{Generating a source catalog}
\label{sec2}
To perform Monte-Carlo simulations of parameter estimation, we start with generating a mock catalog of BBH. We assume circular nonspinning binaries just for simplicity \footnote{Indeed, we confirmed that at 90\% CL, the inclusion of aligned spins change the errors in luminosity distance and sky localization by less than 5\% and 0.01\%, respectively.} and use the PhenomD waveform \cite{Khan:2016PRD}, which is an up-to-date version of inspiral-merger-ringdown waveform for aligned-spinning (nonprecessing) BBH with mass ratio up to 1:18. Then each binary source is described by the following nine physical parameters $\{ {\cal M}, \eta, t_c, \phi_c, d_L, \theta_S, \phi_S, \iota, \psi \}$: intrinsic chirp mass, symmetric mass ratio, time and phase at coalescence, luminosity distance, two angles corresponding to the source sky direction, an angle between the direction of orbital angular momentum and line of sight, and polarization angle. The distribution of each component mass is drawn from the mass-weighted distributions, $m^{-\alpha}$, with $\alpha=2.35$ (Salpeter-type) and $\alpha=1$ (log-flat). Each component mass ranges from $5\,M_{\odot}$ to $100\,M_{\odot}$, but the total mass does not exceed $100\,M_{\odot}$, as assumed in the analysis of \cite{GW170104:detection}. Both of these distributions are still allowed observationally, but the corresponding BBH merger rates are different \cite{GW170104:detection}. For each BBH, we randomly choose the directions of BBH on the sky and its orbital angular momentum. We assume a constant merger rate per unit comoving volume and unit time. This is a conservative assumption because the BBH merger rate in the scenario of isolated field binaries is predicted to increase up to $z\sim 2$ by an order of magnitude \cite{Belczynski:2016Nature}. The fiducial cosmological parameters are set to those from Planck \cite{Planck2015cosmology}, assuming a flat Lambda cold-dark-matter ($\Lambda$CDM) cosmology. We limit source redshifts to $z<0.3$, because as we will see later, well-localized sources that can be used for determination of the Hubble constant are concentrated at low redshifts ($z\lesssim0.1$). The signal-to-noise ratio (SNR) $\rho$ of each BBH is computed from
\begin{equation}
\rho^2 = 4 \sum_{I} \int_{f_{\rm{min}}}^{f_{\rm{max}}} \frac{|\tilde{h}_I (f)|^2}{S_h(f)} df \;, \label{eq1}
\end{equation}
where $\tilde{h}_I$ is the Fourier amplitude of a GW signal in $I$th detector and $S_{h}$ is the noise power spectral density of a detector. The summation in Eq.~(\ref{eq1}) is taken over all detectors under consideration. We consider detector networks composed of aLIGO H1 (H), aLIGO L1 (L), aVIRGO (V), and KAGRA (K), setting locations and orientations to realistic ones. The minimum and maximum frequencies are $f_{\rm min}=30\,{\rm Hz}$ and $f_{\rm max}=10\,{\rm kHz}$, respectively. 

%The noise curves are assumed to be the same as that of aLIGO in \cite{Sathyaprakash:2009xs}. 

We repeat the above procedure and generate 50000 sources up to $z=0.3$ for our source catalog. Then SNR is computed for each source and only sources with $\rho>8$ are kept as observed ones. The parameter probability distributions of the GW sources  are shown in red in Fig.~\ref{fig1}.

%%%%%%%%%%%%%%%%%%%%%%%%%%%%%%%%

\section{Host galaxy identification}
\label{sec3}
We compute parameter estimation errors for each BBH with a Fisher information matrix with the nine parameters for a nonspinning binary. The parameters we are interested in for the purpose of host-galaxy identification are luminosity distance and sky localization area. The sky localization error is computed by
\begin{equation}
\Delta \Omega_{\rm S} \equiv 2 \pi | \sin \theta_{S}| \sqrt{(\Delta \theta_{\rm S})^2(\Delta \phi_{\rm S})^2 - \langle \delta \theta_{\rm S} \delta \phi_{\rm S} \rangle^2 } \;,
\end{equation}
where $\langle \cdots \rangle$ stands for ensemble average and $\Delta \theta_{\rm S} \equiv \langle (\delta \theta_{\rm S})^2 \rangle^{1/2}$ and $\Delta \phi_{\rm S} \equiv \langle (\delta \phi_{\rm S})^2 \rangle^{1/2}$.

\begin{figure*}[t]
\begin{center}
\raisebox{0.7mm}{\includegraphics[width=6cm]{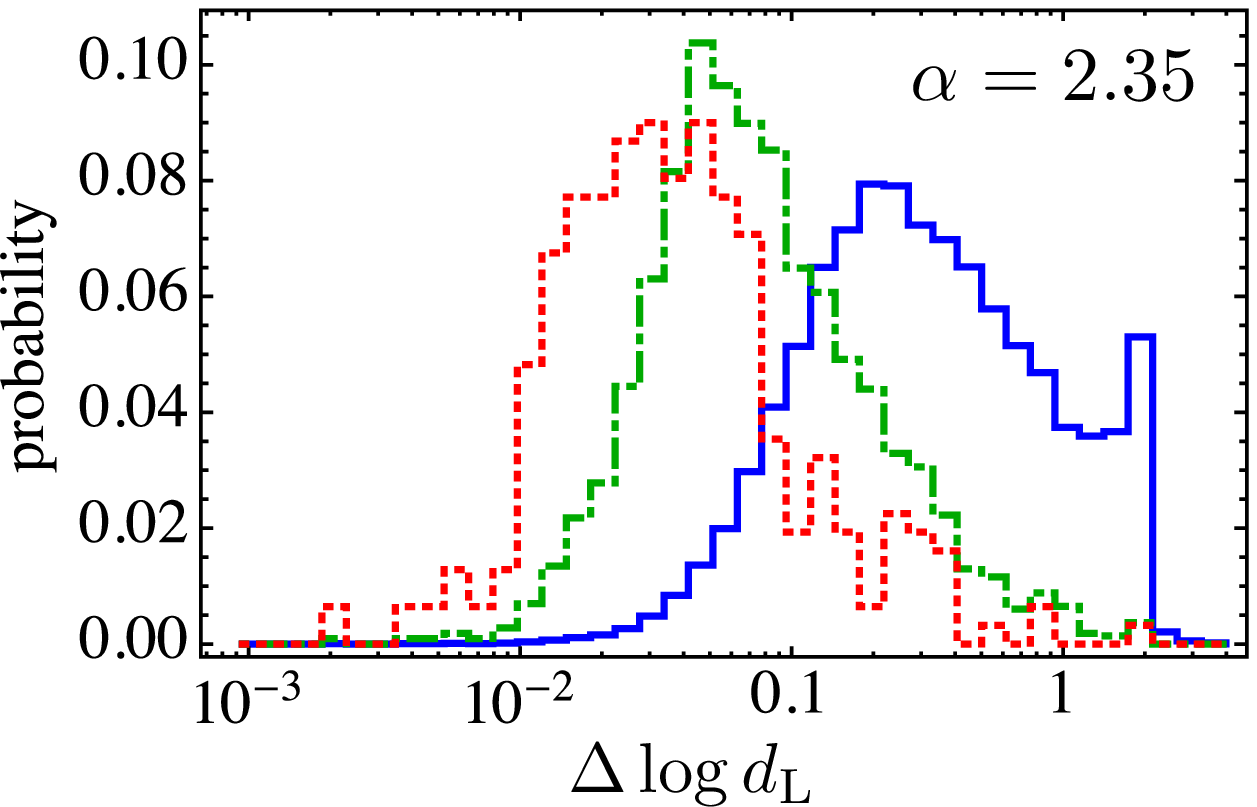}}
\hspace{5mm}
\includegraphics[width=6cm]{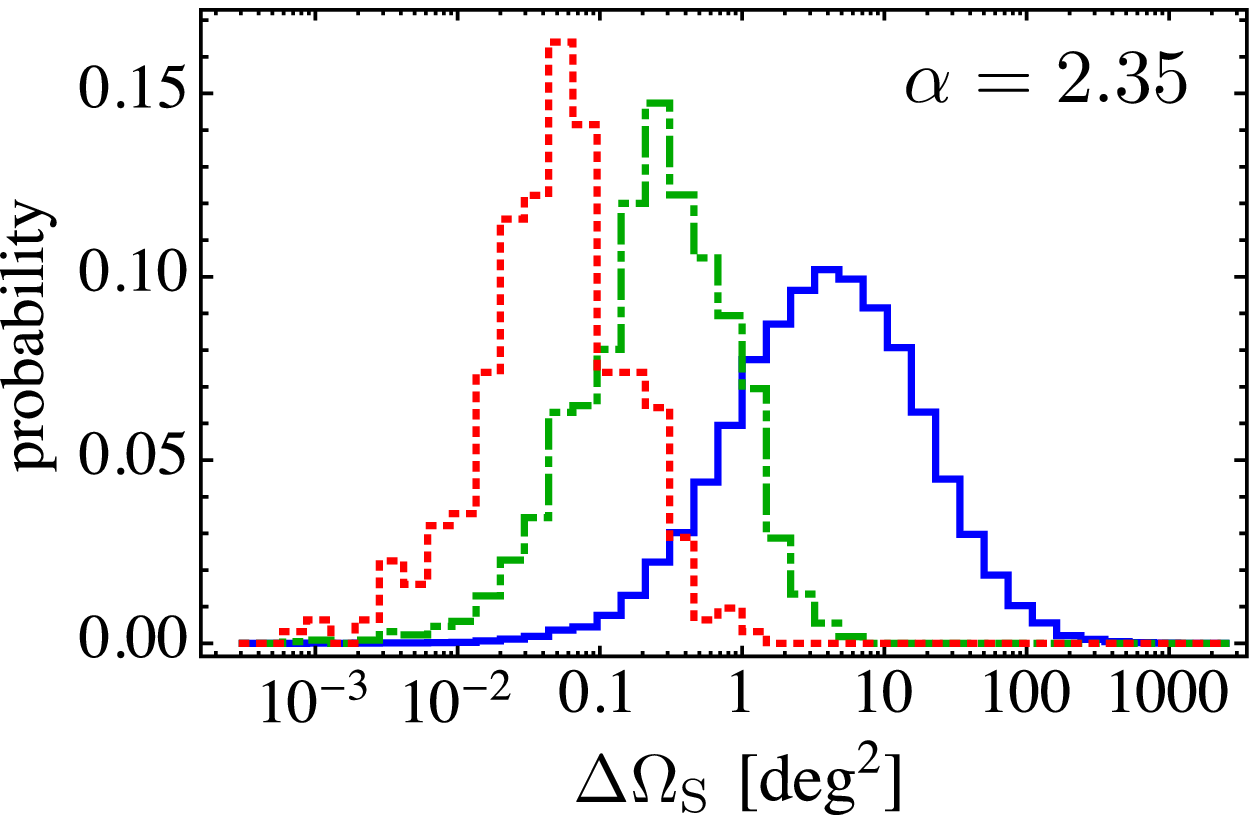}
\raisebox{0.7mm}{\includegraphics[width=6cm]{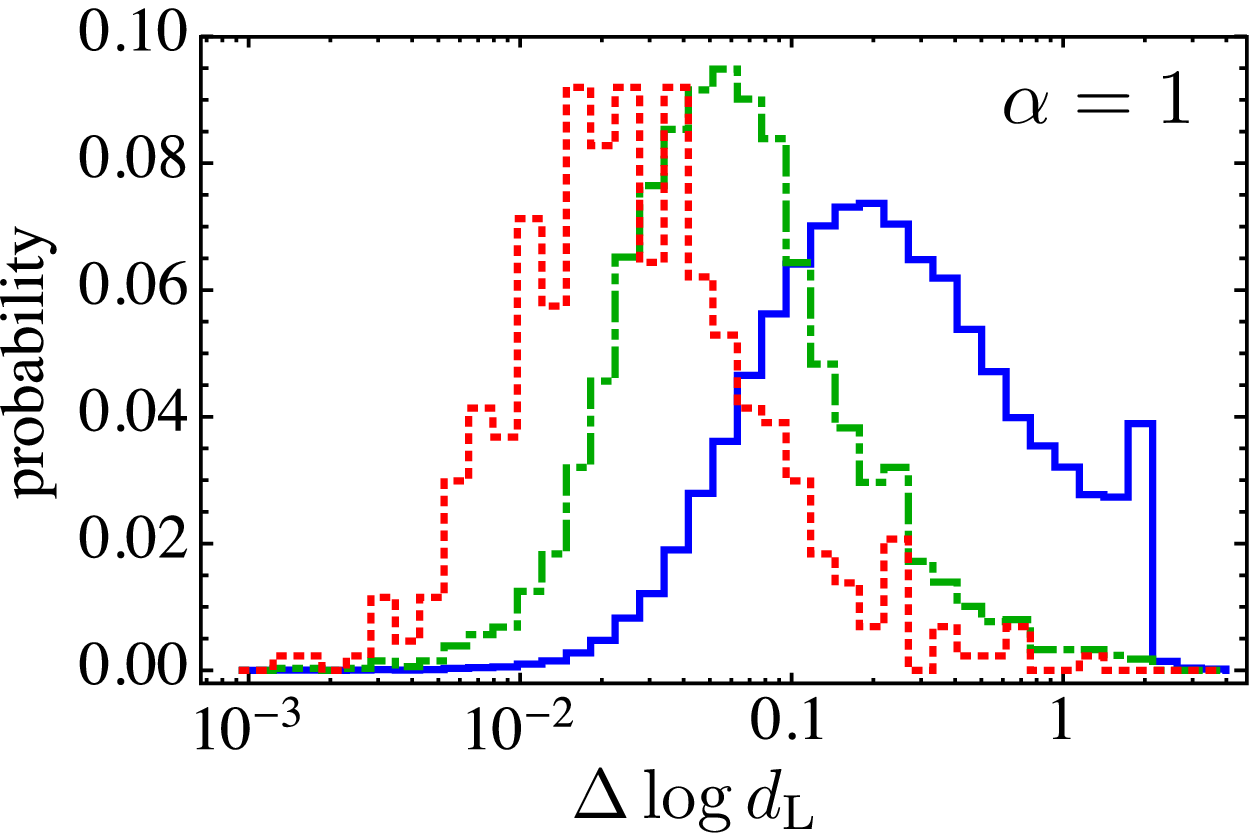}}
\hspace{5mm}
\includegraphics[width=6cm]{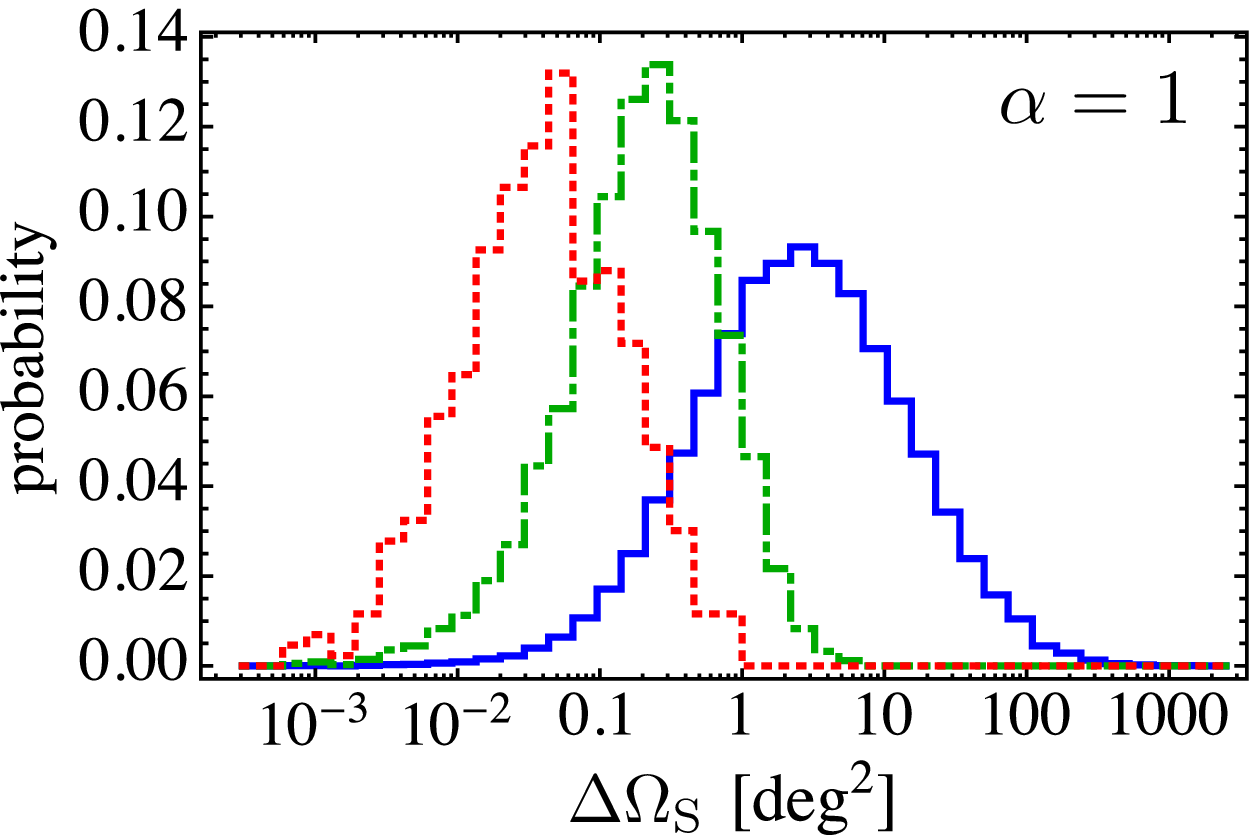}
\caption{Probability distributions of parameter estimation errors for sources from the mass distributions ($\alpha=2.35$ and $\alpha=1$), observed by HLV detector network. The horizontal axes are a relative error in luminosity distance (left) and sky localization error in the unit of square degrees (right). The line colors are the same as Fig.~\ref{fig1}.}
\label{fig2}
\end{center}
\end{figure*}

In Fig.~\ref{fig2}, we show the error probability distributions of luminosity distance and sky localization area. A typical fractional error in luminosity distance is from 0.05 - 2 (undetermined), while a typical sky localization error is 0.1 - 100 ${\rm deg}^2$. On the tails of the distributions, however, there exists a small population of BBH that has significantly smaller errors in the distance and sky localization. Although its fraction is small, non-negligible number of such BBH is observed if the total number of BBH observed is large. It is possible for these golden binaries to uniquely identify a host galaxy in each sky localization volume and obtain a source redshift. 

Assuming that the number density of galaxies is $n_{\rm gal}=0.01\,{\rm Mpc}^{-3}$, which covers roughly 90\% of the total luminosity in B-band \cite{Gehrels:2015ApJ}, we count the number of galaxies $N_{\rm host}$ in sky localization error volume at 90\% CL for each BBH by  
\begin{equation}
N_{\rm host} \equiv n_{\rm gal} \left[ V (d_{L,\rm max}) - V (d_{L,\rm min}) \right] \frac{\Delta \Omega_{\rm S}}{4\pi} \;.
\label{eq:Verr}
\end{equation}
Here $V(d_L)$ is comoving volume of a sphere with radius $d_L$. The maximum and minimum luminosity distances are determined by $d_{L,{\rm max}} = d_{L} (z_{\rm f}) + \Delta d_{L}$ and $d_{L,{\rm min}} = \max \left[ d_L(z_{\rm f}) - \Delta d_{L}, 0 \right]$, where $z_{\rm f}$ is a fiducial source redshift and $\Delta d_{L}$ is a parameter estimation error of luminosity distance. In these conversion between a redshift and luminosity distance, we used fiducial cosmological parameters, which may cause a bias in cosmological parameter estimation, but it is a higher order effect and can be ignored for our purpose to investigate leading-order measurability of the Hubble constant in aLIGO era.

\begin{figure}[t]
\begin{center}
\includegraphics[width=7.5cm]{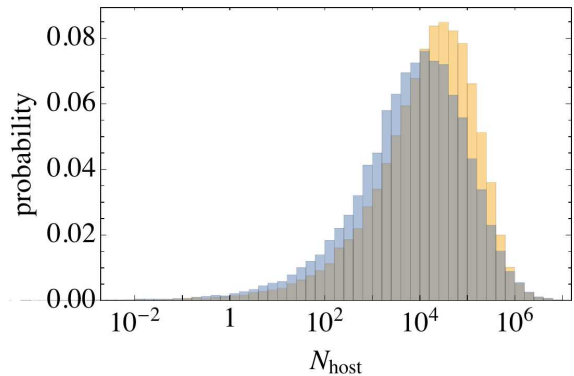}
\caption{Probability distributions of the number of host galaxy candidates for a BBH merger event observed by HLV network. The color shows the mass distributions, $\alpha=2.35$ (orange) and $\alpha=1$ (blue).}
\label{fig3}
\end{center}
\end{figure}

In Fig.~\ref{fig3}, the probability distribution of the number of host galaxy candidates in sky localization volume for each BBH is plotted. Most of BBH has $10^2$ - $10^6$ galaxies in their sky localization volume. However, there exists a small number of BBH that can identify a unique host galaxy. The fractions of these BBH among all BBH observed is 0.74\% for HLV network and 1.4\% for HLVK network in $\alpha=2.35$ case, and 1.0\% for HLV network and 2.2\% for HLVK network in $\alpha=1$ case, as listed in Table~\ref{tab1}. In either case, the success probabilities of host-galaxy identification are rather small, but a large number of sources observed leads to a non-negligible number of host-galaxy identified sources. In Figs.~\ref{fig1} and \ref{fig2}, we plot the probability distributions of these BBH subclasses filtered by $N_{\rm host}<100$ and $N_{\rm host}<1$. As seen from Fig.~\ref{fig1}, chirp masses are almost independent of the number of host galaxy candidates. However, well-localized sources are those at rather low redshifts and consequently with high SNR. Indeed, in Fig.~\ref{fig2}, these BBH have much smaller distance and sky localization errors. Therefore, our statistical study reveals that a small number of BBH at significantly low redshifts enables us to obtain their redshifts by identifying their host galaxies from a spectroscopic follow-up observation or just referring to a nearly complete galaxy catalog at $z \lesssim 0.1$. This conclusion has also been reached in a recent work on 3D error volume and host galaxy identification by Chen and Holz \cite{Chen:2016tys}. We note that the number of sources with $N_{\rm host}<1$ for HLVK network is roughly twice of HLV network. This simply results from a sky localization error twice better due to extending a detector network.

\begin{table*}[t]
\begin{center}
\begin{tabular}{|lcccc|}
\hline
$\alpha=2.35$ case & & \multicolumn{3}{c|}{event rate $[{\rm{yr}}^{-1}]$} \\ \cline{3-5}
 source selection & source catalog \quad \quad & $\;220\,{\rm Gpc}^{-3}\,{\rm yr}^{-1}\;$ & $\;130\,{\rm Gpc}^{-3}\,{\rm yr}^{-1}\;$ & $\;40\,{\rm Gpc}^{-3}\,{\rm yr}^{-1}\;$ \\
\hline
BBH merger ($z<0.3$) & 50000 & 1956 & 1156 & 356 \\
HLV ($\rho >$8, $z<0.3$) & 47982 & 1877 & 1109 & 341 \\
HLV ($\rho >$8, $z<0.3$, $N_{\rm host} <1$) & 353 & 14 & 8 & 3 \\
HLVK ($\rho >$8, $z<0.3$) & 49361 & 1931 & 1141 & 351 \\
HLVK ($\rho >$8, $z<0.3$, $N_{\rm host} <1$) & 696 & 27 & 16 & 5 \\
\hline 
\end{tabular}
\end{center}
\caption{Expected number of BBH observed by HLV and HLVK when $\alpha=2.35$ mass distribution is assumed. The selected BBH merger rates correspond to the current maximum, intermediate, and minimum ones from the aLIGO observation \cite{GW170104:detection}.  The numbers in specific cases of BBH merger rates are scaled from those obtained for the source catalog, containing Poissonian errors up to roughly 5.3\% and 3.8\% for HLV and HLVK, respectively.}
\label{tab1}
\end{table*}

\begin{table*}[t]
\begin{center}
\begin{tabular}{|lcccc|}
\hline
$\alpha=1$ case & & \multicolumn{3}{c|}{event rate $[{\rm{yr}}^{-1}]$} \\ \cline{3-5}
 source selection & source catalog \quad \quad & $\;70\,{\rm Gpc}^{-3}\,{\rm yr}^{-1}\;$ & $\;30\,{\rm Gpc}^{-3}\,{\rm yr}^{-1}\;$ & $\;10\,{\rm Gpc}^{-3}\,{\rm yr}^{-1}\;$ \\
\hline
BBH merger ($z<0.3$) & 50000 & 622 & 267 & 89 \\
HLV ($\rho >$8, $z<0.3$) & 49030 & 610 & 262 & 87 \\
HLV ($\rho >$8, $z<0.3$, $N_{\rm host} <1$) & 495 & 6 & 3 & 1 \\
HLVK ($\rho >$8, $z<0.3$) & 49721 & 619 & 265 & 88 \\
HLVK ($\rho >$8, $z<0.3$, $N_{\rm host} <1$) & 1083 & 14 & 6 & 2 \\
\hline 
\end{tabular}
\end{center}
\caption{Same as Table~\ref{tab1}, but the mass distribution is different, $\alpha=1$ here. Correspondingly, the selected merger rates are different.}
\label{tab2}
\end{table*}

%%%%%%%%%%%%%%%%%%%%%%%%%%%%%%%%

\section{Measurement of the Hubble constant} 
\label{sec4}
With golden binaries whose redshifts are known from host galaxies, we estimate a measurement error of the Hubble constant with a Fisher matrix. In the flat $\Lambda$CDM cosmology, the cosmic expansion history is described by two parameters: Hubble constant $H_0$ and matter energy density $\Omega_{\rm m}$. Since GW golden binaries at low redshifts are not sensitive to $\Omega_{\rm m}$, which plays a role only at high redshifts, we adopt a Gaussian prior $\Delta \Omega_{\rm m}=0.013$ from the CMB observation by Planck \cite{Planck2015cosmology}. There are two systematic errors that can contribute to a luminosity distance measurement \cite{Bertacca:2017vod}: gravitational lensing and galaxy peculiar velocity. The former directly changes apparent luminosity distance by magnifying/demagnifying GW amplitude, but it is negligible because the golden binaries are at low redshifts. The latter affects a measured redshift via the Doppler shift in a spectroscopic measurement of galaxy and indirectly contributes to an error in luminosity distance \cite{Gordon2007PRL,Nishizawa:2010xx}. The systematic error due to the peculiar velocity $\sigma_{\rm pv}$ is more important at lower redshifts. The total error in luminosity distance is defined as 
\begin{equation}
\sigma_{d_L}^2 (z) = \sigma_{\rm GW}^2(z) + \sigma_{\rm pv}^2(z) \;,
\end{equation}
where $\sigma_{\rm GW}$ is a luminosity distance error purely from a GW observation and $\sigma_{\rm pv}$ is given by
\begin{align}
\sigma_{\rm{pv}}(z) = \left| 1-\frac{(1+z)^2}{H(z)d_L(z)} \right| \sigma_{\rm{v,gal}} \;. \nonumber
\end{align}
We set the radial velocity dispersion of galaxies to $\sigma_{\rm{v,gal}}=300\,$km\,s$^{-1}$.
Then we estimate measurement errors of the cosmological parameters, $H_0$ and $\Omega_{\rm m}$, from the Fisher matrix:
\begin{equation}
\Gamma_{ab} = \sum_i \frac{\partial_a d_L (z_i) \partial_b d_L(z_i)}{\sigma_{d_L}^2(z_i)} \;, 
\end{equation}
where $i$ runs over all redshift-identified GW sources and $\partial_a$ is a derivative with respect to $H_0$ or $\Omega_{\rm m}$.

\begin{figure}[t]
\begin{center}
\includegraphics[width=7.5cm]{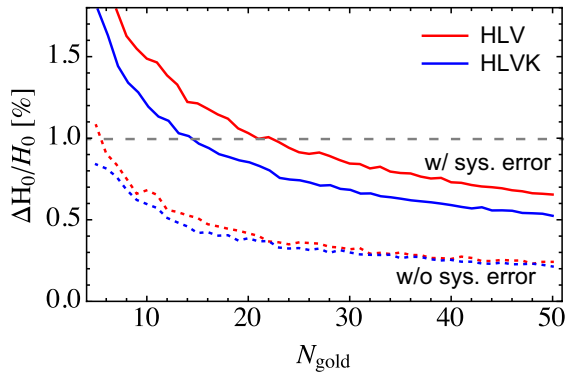}
\caption{Measurement precision of the Hubble constant as a function of the observed number of golden BBH in the case of $\alpha=2.35$ mass distribution. The detector networks are HLV (red) and HLVK (blue). The solid and dotted lines are with/without the systematic error from peculiar velocity. The horizontal line shows 1\% precision.}
\label{fig4}
\end{center}
\end{figure}

Let us denote the observed number of BBH with $N_{\rm host}<1$ by $N_{\rm gold}$. For each $N_{\rm gold}$ up to 50, we take 100 sets of $N_{\rm gold}$ binaries randomly sampled from our source catalog and average the cosmological parameter errors over the realizations. The average $\Delta H_0/H_0$ is shown as a function of $N_{\rm gold}$ in Fig.~\ref{fig4}. As the number of observed BBH increases, the fractional error of $H_0$ decreases down to $1.5\%$, $0.85\%$, and $0.65\%$ with HLV network and $1.2\%$, $0.69\%$, and $0.52\%$ with HLVK network in the presence of the systematic error when using 10, 30 and 50 BBH, respectively. In the absence of the systematic error, the sensitivities of HLV and HLVK networks are almost same. This is just because high-SNR events are always detected with both networks and the fractional error in luminosity distance has almost no difference between them except for statistical fluctuations if SNR is fixed. However, in the presence of the systematic error from peculiar velocity, HLVK network is slightly more sensitive for the same number of $N_{\rm gold}$. The reason is because HLVK network can detect sources at further distance, where the systematic error is slightly smaller. More importantly, the largest advantage of HLVK network is that the number of golden binaries expected to be observed is nearly twice.

%%%%%%%%%%%%%%%%%%%%%%%%%%%%%%%%
\section{Discussions} 
\label{sec5}
\subsection{Calibration error}
Current data of GW amplitude from aLIGO observations include at most 5\% uncertainty from a calibration error at one-sigma level \cite{GW170104:detection}. This directly affects the measurement precision of luminosity distance and is not averaged away because of a systematic error. Although we did not take into account the calibration error in our analysis, it should be seriously considered and mitigated in the future observations to achieve the potential sensitivity of GW detectors to the Hubble constant. It would be possible to reduce the calibration error to 1\% level with the sophisticated method proposed by Tuyenbayev {\it et al.} \cite{Tuyenbayev:2016CQG}.

\subsection{Computation of sky localization volume}
One of simplifications in our analysis is the definition of sky localization volume in Eq.~(\ref{eq:Verr}), which slightly overestimates the error volume. In a real data analysis, however, the shape of an error volume is much more complicated, because luminosity distance error and sky localization error are correlated in a nontrivial manner. Thus, an error volume is expected to be more like an ellipsoid \cite{Singer2016ApJL}, indicating that the corners of our error volume should be truncated. To estimate an error volume more accurately, we need to go beyond the Fisher matrix analysis, though the other method like fully coherent Bayesian analysis is computationally much intensive.

\subsection{Galaxy clustering and properties}
Another simplification in our analysis is ignorance of galaxy clustering and properties. We assumed that galaxy number density is spatially constant, but in reality, galaxies are more concentrated in denser regions due to gravity and halo bias. This clustering would make more difficult to find a unique host galaxy and reduce $N_{\rm gold}$ available in our analysis. Then it takes more years to reach the same sensitivity to $H_0$. Even if a unique host galaxy is not found, one can treat the redshift distribution of host galaxy candidates in a Bayesian statistical framework as in \cite{MacLeod:2008PRD,Petiteau:2011ApJ,DelPozzo:2012PRD}. This may improve a measurement precision of the Hubble constant by using full information about GW sources, though a further study on possible observational biases due to missing galaxies on a catalog is necessary. On the other hand, if one filters host galaxy candidates by galaxy properties, e.~g. metallicity, it would be easier to find a unique host galaxy. At present we cannot conclude if these associations are true, but the future GW detections would provide more evidences on the properties of host galaxies.

%\subsection{Validity of the use of Fisher information matrix}
%It is well known that parameter errors given by the Fisher information matrix are sometimes overestimated or underestimated when SNR is small because error distribution functions cannot be approximated by Gaussian distributions \cite{Vallisneri:2007ev}. Consequently, it affects the error estimation of Hubble constant. However, since golden binaries that we used for the measurement of the Hubble constant has large SNR ($>100$), the Gaussian approximation of the distribution functions holds well and the use of the Fisher matrix is justified.

%%%%%%%%%%%%%%%%%%%%%%%%%%%%%%%%
\section{Conclusion} 
\label{sec6}
We have considered a measurement of the Hubble constant with stellar-mass BBH. We found that a small number of BBH has significantly small error volume, which enable us to identify a unique host galaxy and then obtain the redshift of BBH from a spectroscopic follow-up observation without preparing a galaxy catalog {\it a priori}. With the golden GW events, we have shown that the Hubble constant can be determined at the precision better than 1\% only if a calibration error is reduced to that level \cite{Tuyenbayev:2016CQG}. Therefore, future GW observations will help resolve a well-known discrepancy problem between cosmological measurements and local measurements.

\begin{acknowledgments}
The author acknowledges Emanuele Berti, Toshiya Namikawa, and Alberto Sesana for fruitful discussions. This work was supported by NSF Grant PHY-1607130 and the H2020-MSCA-RISE- 2015 Grant No.~StronGrHEP-690904. 
\end{acknowledgments}

\bibliography{/Volumes/USB-MEMORY/my-research/bibliography}

\begin{thebibliography}{33}%
\makeatletter
\providecommand \@ifxundefined [1]{%
 \@ifx{#1\undefined}
}%
\providecommand \@ifnum [1]{%
 \ifnum #1\expandafter \@firstoftwo
 \else \expandafter \@secondoftwo
 \fi
}%
\providecommand \@ifx [1]{%
 \ifx #1\expandafter \@firstoftwo
 \else \expandafter \@secondoftwo
 \fi
}%
\providecommand \natexlab [1]{#1}%
\providecommand \enquote  [1]{``#1''}%
\providecommand \bibnamefont  [1]{#1}%
\providecommand \bibfnamefont [1]{#1}%
\providecommand \citenamefont [1]{#1}%
\providecommand \href@noop [0]{\@secondoftwo}%
\providecommand \href [0]{\begingroup \@sanitize@url \@href}%
\providecommand \@href[1]{\@@startlink{#1}\@@href}%
\providecommand \@@href[1]{\endgroup#1\@@endlink}%
\providecommand \@sanitize@url [0]{\catcode `\\12\catcode `\$12\catcode
  `\&12\catcode `\#12\catcode `\^12\catcode `\_12\catcode `\%12\relax}%
\providecommand \@@startlink[1]{}%
\providecommand \@@endlink[0]{}%
\providecommand \url  [0]{\begingroup\@sanitize@url \@url }%
\providecommand \@url [1]{\endgroup\@href {#1}{\urlprefix }}%
\providecommand \urlprefix  [0]{URL }%
\providecommand \Eprint [0]{\href }%
\providecommand \doibase [0]{http://dx.doi.org/}%
\providecommand \selectlanguage [0]{\@gobble}%
\providecommand \bibinfo  [0]{\@secondoftwo}%
\providecommand \bibfield  [0]{\@secondoftwo}%
\providecommand \translation [1]{[#1]}%
\providecommand \BibitemOpen [0]{}%
\providecommand \bibitemStop [0]{}%
\providecommand \bibitemNoStop [0]{.\EOS\space}%
\providecommand \EOS [0]{\spacefactor3000\relax}%
\providecommand \BibitemShut  [1]{\csname bibitem#1\endcsname}%
\let\auto@bib@innerbib\@empty
%</preamble>
\bibitem [{\citenamefont {Ade}\ \emph {et~al.}(2016)\citenamefont {Ade} \emph
  {et~al.}}]{Planck2015cosmology}%
  \BibitemOpen
  \bibfield  {author} {\bibinfo {author} {\bibfnamefont {P.~A.~R.}\
  \bibnamefont {Ade}} \emph {et~al.} (\bibinfo {collaboration} {Planck
  collaboration}),\ }\href {\doibase 10.1051/0004-6361/201525830} {\bibfield
  {journal} {\bibinfo  {journal} {Astron. Astrophys.}\ }\textbf {\bibinfo
  {volume} {594}},\ \bibinfo {pages} {A13} (\bibinfo {year} {2016})},\ \Eprint
  {http://arxiv.org/abs/1502.01589} {arXiv:1502.01589 [astro-ph.CO]}
  \BibitemShut {NoStop}%
%%CITATION = ARXIV:1502.01589;%%
\bibitem [{\citenamefont {Beutler}\ \emph {et~al.}(2011)\citenamefont
  {Beutler}, \citenamefont {Blake}, \citenamefont {Colless}, \citenamefont
  {Jones}, \citenamefont {Staveley-Smith}, \citenamefont {Campbell},
  \citenamefont {Parker}, \citenamefont {Saunders},\ and\ \citenamefont
  {Watson}}]{Beutler:2011MNRAS}%
  \BibitemOpen
  \bibfield  {author} {\bibinfo {author} {\bibfnamefont {F.}~\bibnamefont
  {Beutler}}, \bibinfo {author} {\bibfnamefont {C.}~\bibnamefont {Blake}},
  \bibinfo {author} {\bibfnamefont {M.}~\bibnamefont {Colless}}, \bibinfo
  {author} {\bibfnamefont {D.~H.}\ \bibnamefont {Jones}}, \bibinfo {author}
  {\bibfnamefont {L.}~\bibnamefont {Staveley-Smith}}, \bibinfo {author}
  {\bibfnamefont {L.}~\bibnamefont {Campbell}}, \bibinfo {author}
  {\bibfnamefont {Q.}~\bibnamefont {Parker}}, \bibinfo {author} {\bibfnamefont
  {W.}~\bibnamefont {Saunders}}, \ and\ \bibinfo {author} {\bibfnamefont
  {F.}~\bibnamefont {Watson}},\ }\href {\doibase
  10.1111/j.1365-2966.2011.19250.x} {\bibfield  {journal} {\bibinfo  {journal}
  {Mon. Not. Roy. Astron. Soc.}\ }\textbf {\bibinfo {volume} {416}},\ \bibinfo
  {pages} {3017} (\bibinfo {year} {2011})},\ \Eprint
  {http://arxiv.org/abs/1106.3366} {arXiv:1106.3366 [astro-ph.CO]} \BibitemShut
  {NoStop}%
%%CITATION = ARXIV:1106.3366;%%
\bibitem [{\citenamefont {Riess}\ \emph {et~al.}(2016)\citenamefont {Riess}
  \emph {et~al.}}]{Riess:2016ApJ}%
  \BibitemOpen
  \bibfield  {author} {\bibinfo {author} {\bibfnamefont {A.~G.}\ \bibnamefont
  {Riess}} \emph {et~al.},\ }\href {\doibase 10.3847/0004-637X/826/1/56}
  {\bibfield  {journal} {\bibinfo  {journal} {Astrophys. J.}\ }\textbf
  {\bibinfo {volume} {826}},\ \bibinfo {pages} {56} (\bibinfo {year} {2016})},\
  \Eprint {http://arxiv.org/abs/1604.01424} {arXiv:1604.01424 [astro-ph.CO]}
  \BibitemShut {NoStop}%
%%CITATION = ARXIV:1604.01424;%%
\bibitem [{\citenamefont {{Jackson}}(2015)}]{Jackson:2015LRR}%
  \BibitemOpen
  \bibfield  {author} {\bibinfo {author} {\bibfnamefont {N.}~\bibnamefont
  {{Jackson}}},\ }\href {\doibase 10.1007/lrr-2015-2} {\bibfield  {journal}
  {\bibinfo  {journal} {Living Reviews in Relativity}\ }\textbf {\bibinfo
  {volume} {18}},\ \bibinfo {eid} {2} (\bibinfo {year} {2015})}\BibitemShut
  {NoStop}%
\bibitem [{\citenamefont {{Abbott}}\ \emph
  {et~al.}(2016{\natexlab{a}})\citenamefont {{Abbott}} \emph
  {et~al.}}]{GW150914:detection}%
  \BibitemOpen
  \bibfield  {author} {\bibinfo {author} {\bibfnamefont {B.~P.}\ \bibnamefont
  {{Abbott}}} \emph {et~al.} (\bibinfo {collaboration} {LIGO Scientific
  Collaboration and Virgo Collaboration}),\ }\href {\doibase
  10.1103/PhysRevLett.116.061102} {\bibfield  {journal} {\bibinfo  {journal}
  {Physical Review Letters}\ }\textbf {\bibinfo {volume} {116}},\ \bibinfo
  {eid} {061102} (\bibinfo {year} {2016}{\natexlab{a}})},\ \Eprint
  {http://arxiv.org/abs/1602.03837} {arXiv:1602.03837 [gr-qc]} \BibitemShut
  {NoStop}%
\bibitem [{\citenamefont {{Abbott}}\ \emph
  {et~al.}(2016{\natexlab{b}})\citenamefont {{Abbott}} \emph
  {et~al.}}]{GW151226:detection}%
  \BibitemOpen
  \bibfield  {author} {\bibinfo {author} {\bibfnamefont {B.~P.}\ \bibnamefont
  {{Abbott}}} \emph {et~al.} (\bibinfo {collaboration} {LIGO Scientific
  Collaboration and Virgo Collaboration}),\ }\href {\doibase
  10.1103/PhysRevLett.116.241103} {\bibfield  {journal} {\bibinfo  {journal}
  {Physical Review Letters}\ }\textbf {\bibinfo {volume} {116}},\ \bibinfo
  {eid} {241103} (\bibinfo {year} {2016}{\natexlab{b}})},\ \Eprint
  {http://arxiv.org/abs/1606.04855} {arXiv:1606.04855 [gr-qc]} \BibitemShut
  {NoStop}%
\bibitem [{\citenamefont {Abbott}\ \emph {et~al.}(2017)\citenamefont {Abbott}
  \emph {et~al.}}]{GW170104:detection}%
  \BibitemOpen
  \bibfield  {author} {\bibinfo {author} {\bibfnamefont {B.~P.}\ \bibnamefont
  {Abbott}} \emph {et~al.} (\bibinfo {collaboration} {LIGO Scientific
  Collaboration}),\ }\href {\doibase 10.1103/PhysRevLett.118.221101} {\bibfield
   {journal} {\bibinfo  {journal} {Phys. Rev. Lett.}\ }\textbf {\bibinfo
  {volume} {118}},\ \bibinfo {pages} {221101} (\bibinfo {year} {2017})},\
  \Eprint {http://arxiv.org/abs/1706.01812} {arXiv:1706.01812 [gr-qc]}
  \BibitemShut {NoStop}%
%%CITATION = ARXIV:1706.01812;%%
\bibitem [{\citenamefont {Belczynski}\ \emph {et~al.}(2010)\citenamefont
  {Belczynski}, \citenamefont {Dominik}, \citenamefont {Bulik}, \citenamefont
  {O'Shaughnessy}, \citenamefont {Fryer},\ and\ \citenamefont
  {Holz}}]{Belczynski:2010ApJL}%
  \BibitemOpen
  \bibfield  {author} {\bibinfo {author} {\bibfnamefont {K.}~\bibnamefont
  {Belczynski}}, \bibinfo {author} {\bibfnamefont {M.}~\bibnamefont {Dominik}},
  \bibinfo {author} {\bibfnamefont {T.}~\bibnamefont {Bulik}}, \bibinfo
  {author} {\bibfnamefont {R.}~\bibnamefont {O'Shaughnessy}}, \bibinfo {author}
  {\bibfnamefont {C.~L.}\ \bibnamefont {Fryer}}, \ and\ \bibinfo {author}
  {\bibfnamefont {D.~E.}\ \bibnamefont {Holz}},\ }\href {\doibase
  10.1088/2041-8205/715/2/L138} {\bibfield  {journal} {\bibinfo  {journal}
  {Astrophys. J.}\ }\textbf {\bibinfo {volume} {715}},\ \bibinfo {pages} {L138}
  (\bibinfo {year} {2010})},\ \Eprint {http://arxiv.org/abs/1004.0386}
  {arXiv:1004.0386 [astro-ph.HE]} \BibitemShut {NoStop}%
%%CITATION = ARXIV:1004.0386;%%
\bibitem [{\citenamefont {Belczynski}\ \emph
  {et~al.}(2016{\natexlab{a}})\citenamefont {Belczynski}, \citenamefont
  {Repetto}, \citenamefont {Holz}, \citenamefont {O'Shaughnessy}, \citenamefont
  {Bulik}, \citenamefont {Berti}, \citenamefont {Fryer},\ and\ \citenamefont
  {Dominik}}]{Belczynski:2015ApJ}%
  \BibitemOpen
  \bibfield  {author} {\bibinfo {author} {\bibfnamefont {K.}~\bibnamefont
  {Belczynski}}, \bibinfo {author} {\bibfnamefont {S.}~\bibnamefont {Repetto}},
  \bibinfo {author} {\bibfnamefont {D.~E.}\ \bibnamefont {Holz}}, \bibinfo
  {author} {\bibfnamefont {R.}~\bibnamefont {O'Shaughnessy}}, \bibinfo {author}
  {\bibfnamefont {T.}~\bibnamefont {Bulik}}, \bibinfo {author} {\bibfnamefont
  {E.}~\bibnamefont {Berti}}, \bibinfo {author} {\bibfnamefont
  {C.}~\bibnamefont {Fryer}}, \ and\ \bibinfo {author} {\bibfnamefont
  {M.}~\bibnamefont {Dominik}},\ }\href {\doibase 10.3847/0004-637X/819/2/108}
  {\bibfield  {journal} {\bibinfo  {journal} {Astrophys. J.}\ }\textbf
  {\bibinfo {volume} {819}},\ \bibinfo {pages} {108} (\bibinfo {year}
  {2016}{\natexlab{a}})},\ \Eprint {http://arxiv.org/abs/1510.04615}
  {arXiv:1510.04615 [astro-ph.HE]} \BibitemShut {NoStop}%
%%CITATION = ARXIV:1510.04615;%%
\bibitem [{\citenamefont {{Schutz}}(1986)}]{Schutz:1986Nature}%
  \BibitemOpen
  \bibfield  {author} {\bibinfo {author} {\bibfnamefont {B.~F.}\ \bibnamefont
  {{Schutz}}},\ }\href {\doibase 10.1038/323310a0} {\bibfield  {journal}
  {\bibinfo  {journal} {\nat}\ }\textbf {\bibinfo {volume} {323}},\ \bibinfo
  {pages} {310} (\bibinfo {year} {1986})}\BibitemShut {NoStop}%
\bibitem [{\citenamefont {{Holz}}\ and\ \citenamefont
  {{Hughes}}(2005)}]{Holz:2005ApJ}%
  \BibitemOpen
  \bibfield  {author} {\bibinfo {author} {\bibfnamefont {D.~E.}\ \bibnamefont
  {{Holz}}}\ and\ \bibinfo {author} {\bibfnamefont {S.~A.}\ \bibnamefont
  {{Hughes}}},\ }\href {\doibase 10.1086/431341} {\bibfield  {journal}
  {\bibinfo  {journal} {\apj}\ }\textbf {\bibinfo {volume} {629}},\ \bibinfo
  {pages} {15} (\bibinfo {year} {2005})},\ \Eprint
  {http://arxiv.org/abs/astro-ph/0504616} {astro-ph/0504616} \BibitemShut
  {NoStop}%
\bibitem [{\citenamefont {Sathyaprakash}\ \emph {et~al.}(2010)\citenamefont
  {Sathyaprakash}, \citenamefont {Schutz},\ and\ \citenamefont {Van
  Den~Broeck}}]{Sathyaprakash:2009xt}%
  \BibitemOpen
  \bibfield  {author} {\bibinfo {author} {\bibfnamefont {B.}~\bibnamefont
  {Sathyaprakash}}, \bibinfo {author} {\bibfnamefont {B.}~\bibnamefont
  {Schutz}}, \ and\ \bibinfo {author} {\bibfnamefont {C.}~\bibnamefont {Van
  Den~Broeck}},\ }\href {\doibase 10.1088/0264-9381/27/21/215006} {\bibfield
  {journal} {\bibinfo  {journal} {Class.Quant.Grav.}\ }\textbf {\bibinfo
  {volume} {27}},\ \bibinfo {pages} {215006} (\bibinfo {year} {2010})},\
  \Eprint {http://arxiv.org/abs/0906.4151} {arXiv:0906.4151 [astro-ph.CO]}
  \BibitemShut {NoStop}%
%%CITATION = ARXIV:0906.4151;%%
\bibitem [{\citenamefont {Nissanke}\ \emph {et~al.}(2013)\citenamefont
  {Nissanke}, \citenamefont {Holz}, \citenamefont {Dalal}, \citenamefont
  {Hughes}, \citenamefont {Sievers},\ and\ \citenamefont
  {Hirata}}]{Nissanke:2013fka}%
  \BibitemOpen
  \bibfield  {author} {\bibinfo {author} {\bibfnamefont {S.}~\bibnamefont
  {Nissanke}}, \bibinfo {author} {\bibfnamefont {D.~E.}\ \bibnamefont {Holz}},
  \bibinfo {author} {\bibfnamefont {N.}~\bibnamefont {Dalal}}, \bibinfo
  {author} {\bibfnamefont {S.~A.}\ \bibnamefont {Hughes}}, \bibinfo {author}
  {\bibfnamefont {J.~L.}\ \bibnamefont {Sievers}}, \ and\ \bibinfo {author}
  {\bibfnamefont {C.~M.}\ \bibnamefont {Hirata}},\ }\href@noop {} {\  (\bibinfo
  {year} {2013})},\ \Eprint {http://arxiv.org/abs/1307.2638} {arXiv:1307.2638
  [astro-ph.CO]} \BibitemShut {NoStop}%
%%CITATION = ARXIV:1307.2638;%%
\bibitem [{\citenamefont {Tamanini}\ \emph {et~al.}(2016)\citenamefont
  {Tamanini}, \citenamefont {Caprini}, \citenamefont {Barausse}, \citenamefont
  {Sesana}, \citenamefont {Klein},\ and\ \citenamefont
  {Petiteau}}]{Tamanini:2016JCAP}%
  \BibitemOpen
  \bibfield  {author} {\bibinfo {author} {\bibfnamefont {N.}~\bibnamefont
  {Tamanini}}, \bibinfo {author} {\bibfnamefont {C.}~\bibnamefont {Caprini}},
  \bibinfo {author} {\bibfnamefont {E.}~\bibnamefont {Barausse}}, \bibinfo
  {author} {\bibfnamefont {A.}~\bibnamefont {Sesana}}, \bibinfo {author}
  {\bibfnamefont {A.}~\bibnamefont {Klein}}, \ and\ \bibinfo {author}
  {\bibfnamefont {A.}~\bibnamefont {Petiteau}},\ }\href {\doibase
  10.1088/1475-7516/2016/04/002} {\bibfield  {journal} {\bibinfo  {journal}
  {JCAP}\ }\textbf {\bibinfo {volume} {1604}},\ \bibinfo {pages} {002}
  (\bibinfo {year} {2016})},\ \Eprint {http://arxiv.org/abs/1601.07112}
  {arXiv:1601.07112 [astro-ph.CO]} \BibitemShut {NoStop}%
%%CITATION = ARXIV:1601.07112;%%
\bibitem [{\citenamefont {Yonetoku}\ \emph {et~al.}(2014)\citenamefont
  {Yonetoku}, \citenamefont {Nakamura}, \citenamefont {Takahashi},\ and\
  \citenamefont {Toyanago}}]{Yonetoku:2014fua}%
  \BibitemOpen
  \bibfield  {author} {\bibinfo {author} {\bibfnamefont {D.}~\bibnamefont
  {Yonetoku}}, \bibinfo {author} {\bibfnamefont {T.}~\bibnamefont {Nakamura}},
  \bibinfo {author} {\bibfnamefont {K.}~\bibnamefont {Takahashi}}, \ and\
  \bibinfo {author} {\bibfnamefont {A.}~\bibnamefont {Toyanago}},\ }\href
  {\doibase 10.1088/0004-637X/789/1/65} {\bibfield  {journal} {\bibinfo
  {journal} {Astrophys.J.}\ }\textbf {\bibinfo {volume} {789}},\ \bibinfo
  {pages} {65} (\bibinfo {year} {2014})},\ \Eprint
  {http://arxiv.org/abs/1402.5463} {arXiv:1402.5463 [astro-ph.HE]} \BibitemShut
  {NoStop}%
%%CITATION = ARXIV:1402.5463;%%
\bibitem [{\citenamefont {Messenger}\ and\ \citenamefont
  {Read}(2012)}]{Messenger:2011gi}%
  \BibitemOpen
  \bibfield  {author} {\bibinfo {author} {\bibfnamefont {C.}~\bibnamefont
  {Messenger}}\ and\ \bibinfo {author} {\bibfnamefont {J.}~\bibnamefont
  {Read}},\ }\href@noop {} {\bibfield  {journal} {\bibinfo  {journal} {Phys.
  Rev. Lett.}\ }\textbf {\bibinfo {volume} {108}},\ \bibinfo {pages} {091101}
  (\bibinfo {year} {2012})},\ \Eprint {http://arxiv.org/abs/1107.5725}
  {arXiv:1107.5725 [gr-qc]} \BibitemShut {NoStop}%
%%CITATION = ARXIV:1107.5725;%%
\bibitem [{\citenamefont {Taylor}\ \emph {et~al.}(2012)\citenamefont {Taylor},
  \citenamefont {Gair},\ and\ \citenamefont {Mandel}}]{Taylor:2011fs}%
  \BibitemOpen
  \bibfield  {author} {\bibinfo {author} {\bibfnamefont {S.~R.}\ \bibnamefont
  {Taylor}}, \bibinfo {author} {\bibfnamefont {J.~R.}\ \bibnamefont {Gair}}, \
  and\ \bibinfo {author} {\bibfnamefont {I.}~\bibnamefont {Mandel}},\
  }\href@noop {} {\bibfield  {journal} {\bibinfo  {journal} {Phys. Rev.}\
  }\textbf {\bibinfo {volume} {D85}},\ \bibinfo {pages} {023535} (\bibinfo
  {year} {2012})},\ \Eprint {http://arxiv.org/abs/1108.5161} {arXiv:1108.5161
  [gr-qc]} \BibitemShut {NoStop}%
%%CITATION = ARXIV:1108.5161;%%
\bibitem [{\citenamefont {{MacLeod}}\ and\ \citenamefont
  {{Hogan}}(2008)}]{MacLeod:2008PRD}%
  \BibitemOpen
  \bibfield  {author} {\bibinfo {author} {\bibfnamefont {C.~L.}\ \bibnamefont
  {{MacLeod}}}\ and\ \bibinfo {author} {\bibfnamefont {C.~J.}\ \bibnamefont
  {{Hogan}}},\ }\href {\doibase 10.1103/PhysRevD.77.043512} {\bibfield
  {journal} {\bibinfo  {journal} {\prd}\ }\textbf {\bibinfo {volume} {77}},\
  \bibinfo {eid} {043512} (\bibinfo {year} {2008})},\ \Eprint
  {http://arxiv.org/abs/0712.0618} {arXiv:0712.0618} \BibitemShut {NoStop}%
\bibitem [{\citenamefont {{Petiteau}}\ \emph {et~al.}(2011)\citenamefont
  {{Petiteau}}, \citenamefont {{Babak}},\ and\ \citenamefont
  {{Sesana}}}]{Petiteau:2011ApJ}%
  \BibitemOpen
  \bibfield  {author} {\bibinfo {author} {\bibfnamefont {A.}~\bibnamefont
  {{Petiteau}}}, \bibinfo {author} {\bibfnamefont {S.}~\bibnamefont {{Babak}}},
  \ and\ \bibinfo {author} {\bibfnamefont {A.}~\bibnamefont {{Sesana}}},\
  }\href {\doibase 10.1088/0004-637X/732/2/82} {\bibfield  {journal} {\bibinfo
  {journal} {\apj}\ }\textbf {\bibinfo {volume} {732}},\ \bibinfo {eid} {82}
  (\bibinfo {year} {2011})},\ \Eprint {http://arxiv.org/abs/1102.0769}
  {arXiv:1102.0769} \BibitemShut {NoStop}%
\bibitem [{\citenamefont {Del~Pozzo}(2012)}]{DelPozzo:2012PRD}%
  \BibitemOpen
  \bibfield  {author} {\bibinfo {author} {\bibfnamefont {W.}~\bibnamefont
  {Del~Pozzo}},\ }\href {\doibase 10.1103/PhysRevD.86.043011} {\bibfield
  {journal} {\bibinfo  {journal} {Phys. Rev.}\ }\textbf {\bibinfo {volume}
  {D86}},\ \bibinfo {pages} {043011} (\bibinfo {year} {2012})},\ \Eprint
  {http://arxiv.org/abs/1108.1317} {arXiv:1108.1317 [astro-ph.CO]} \BibitemShut
  {NoStop}%
%%CITATION = ARXIV:1108.1317;%%
\bibitem [{\citenamefont {Gehrels}\ \emph {et~al.}(2016)\citenamefont
  {Gehrels}, \citenamefont {Cannizzo}, \citenamefont {Kanner}, \citenamefont
  {Kasliwal}, \citenamefont {Nissanke},\ and\ \citenamefont
  {Singer}}]{Gehrels:2015ApJ}%
  \BibitemOpen
  \bibfield  {author} {\bibinfo {author} {\bibfnamefont {N.}~\bibnamefont
  {Gehrels}}, \bibinfo {author} {\bibfnamefont {J.~K.}\ \bibnamefont
  {Cannizzo}}, \bibinfo {author} {\bibfnamefont {J.}~\bibnamefont {Kanner}},
  \bibinfo {author} {\bibfnamefont {M.~M.}\ \bibnamefont {Kasliwal}}, \bibinfo
  {author} {\bibfnamefont {S.}~\bibnamefont {Nissanke}}, \ and\ \bibinfo
  {author} {\bibfnamefont {L.~P.}\ \bibnamefont {Singer}},\ }\href {\doibase
  10.3847/0004-637X/820/2/136} {\bibfield  {journal} {\bibinfo  {journal}
  {Astrophys. J.}\ }\textbf {\bibinfo {volume} {820}},\ \bibinfo {pages} {136}
  (\bibinfo {year} {2016})},\ \Eprint {http://arxiv.org/abs/1508.03608}
  {arXiv:1508.03608 [astro-ph.HE]} \BibitemShut {NoStop}%
%%CITATION = ARXIV:1508.03608;%%
\bibitem [{\citenamefont {{Namikawa}}\ \emph
  {et~al.}(2016{\natexlab{a}})\citenamefont {{Namikawa}}, \citenamefont
  {{Nishizawa}},\ and\ \citenamefont {{Taruya}}}]{Namikawa:2016PRL}%
  \BibitemOpen
  \bibfield  {author} {\bibinfo {author} {\bibfnamefont {T.}~\bibnamefont
  {{Namikawa}}}, \bibinfo {author} {\bibfnamefont {A.}~\bibnamefont
  {{Nishizawa}}}, \ and\ \bibinfo {author} {\bibfnamefont {A.}~\bibnamefont
  {{Taruya}}},\ }\href {\doibase 10.1103/PhysRevLett.116.121302} {\bibfield
  {journal} {\bibinfo  {journal} {Physical Review Letters}\ }\textbf {\bibinfo
  {volume} {116}},\ \bibinfo {eid} {121302} (\bibinfo {year}
  {2016}{\natexlab{a}})},\ \Eprint {http://arxiv.org/abs/1511.04638}
  {arXiv:1511.04638} \BibitemShut {NoStop}%
\bibitem [{\citenamefont {{Namikawa}}\ \emph
  {et~al.}(2016{\natexlab{b}})\citenamefont {{Namikawa}}, \citenamefont
  {{Nishizawa}},\ and\ \citenamefont {{Taruya}}}]{Namikawa:2016PRD}%
  \BibitemOpen
  \bibfield  {author} {\bibinfo {author} {\bibfnamefont {T.}~\bibnamefont
  {{Namikawa}}}, \bibinfo {author} {\bibfnamefont {A.}~\bibnamefont
  {{Nishizawa}}}, \ and\ \bibinfo {author} {\bibfnamefont {A.}~\bibnamefont
  {{Taruya}}},\ }\href {\doibase 10.1103/PhysRevD.94.024013} {\bibfield
  {journal} {\bibinfo  {journal} {\prd}\ }\textbf {\bibinfo {volume} {94}},\
  \bibinfo {eid} {024013} (\bibinfo {year} {2016}{\natexlab{b}})},\ \Eprint
  {http://arxiv.org/abs/1603.08072} {arXiv:1603.08072} \BibitemShut {NoStop}%
\bibitem [{\citenamefont {{Oguri}}(2016)}]{Oguri:2016PRD}%
  \BibitemOpen
  \bibfield  {author} {\bibinfo {author} {\bibfnamefont {M.}~\bibnamefont
  {{Oguri}}},\ }\href {\doibase 10.1103/PhysRevD.93.083511} {\bibfield
  {journal} {\bibinfo  {journal} {\prd}\ }\textbf {\bibinfo {volume} {93}},\
  \bibinfo {eid} {083511} (\bibinfo {year} {2016})},\ \Eprint
  {http://arxiv.org/abs/1603.02356} {arXiv:1603.02356} \BibitemShut {NoStop}%
\bibitem [{Note1()}]{Note1}%
  \BibitemOpen
  \bibinfo {note} {Indeed, we confirmed that at 90\% CL, the inclusion of
  aligned spins change the errors in luminosity distance and sky localization
  by less than 5\% and 0.01\%, respectively.}\BibitemShut {Stop}%
\bibitem [{\citenamefont {{Khan}}\ \emph {et~al.}(2016)\citenamefont {{Khan}},
  \citenamefont {{Husa}}, \citenamefont {{Hannam}}, \citenamefont {{Ohme}},
  \citenamefont {{P{\"u}rrer}}, \citenamefont {{Forteza}},\ and\ \citenamefont
  {{Boh{\'e}}}}]{Khan:2016PRD}%
  \BibitemOpen
  \bibfield  {author} {\bibinfo {author} {\bibfnamefont {S.}~\bibnamefont
  {{Khan}}}, \bibinfo {author} {\bibfnamefont {S.}~\bibnamefont {{Husa}}},
  \bibinfo {author} {\bibfnamefont {M.}~\bibnamefont {{Hannam}}}, \bibinfo
  {author} {\bibfnamefont {F.}~\bibnamefont {{Ohme}}}, \bibinfo {author}
  {\bibfnamefont {M.}~\bibnamefont {{P{\"u}rrer}}}, \bibinfo {author}
  {\bibfnamefont {X.~J.}\ \bibnamefont {{Forteza}}}, \ and\ \bibinfo {author}
  {\bibfnamefont {A.}~\bibnamefont {{Boh{\'e}}}},\ }\href {\doibase
  10.1103/PhysRevD.93.044007} {\bibfield  {journal} {\bibinfo  {journal}
  {\prd}\ }\textbf {\bibinfo {volume} {93}},\ \bibinfo {eid} {044007} (\bibinfo
  {year} {2016})},\ \Eprint {http://arxiv.org/abs/1508.07253} {arXiv:1508.07253
  [gr-qc]} \BibitemShut {NoStop}%
\bibitem [{\citenamefont {Belczynski}\ \emph
  {et~al.}(2016{\natexlab{b}})\citenamefont {Belczynski}, \citenamefont {Holz},
  \citenamefont {Bulik},\ and\ \citenamefont
  {O'Shaughnessy}}]{Belczynski:2016Nature}%
  \BibitemOpen
  \bibfield  {author} {\bibinfo {author} {\bibfnamefont {K.}~\bibnamefont
  {Belczynski}}, \bibinfo {author} {\bibfnamefont {D.~E.}\ \bibnamefont
  {Holz}}, \bibinfo {author} {\bibfnamefont {T.}~\bibnamefont {Bulik}}, \ and\
  \bibinfo {author} {\bibfnamefont {R.}~\bibnamefont {O'Shaughnessy}},\ }\href
  {\doibase 10.1038/nature18322} {\bibfield  {journal} {\bibinfo  {journal}
  {Nature}\ }\textbf {\bibinfo {volume} {534}},\ \bibinfo {pages} {512}
  (\bibinfo {year} {2016}{\natexlab{b}})},\ \Eprint
  {http://arxiv.org/abs/1602.04531} {arXiv:1602.04531 [astro-ph.HE]}
  \BibitemShut {NoStop}%
%%CITATION = ARXIV:1602.04531;%%
\bibitem [{\citenamefont {Chen}\ and\ \citenamefont
  {Holz}(2016)}]{Chen:2016tys}%
  \BibitemOpen
  \bibfield  {author} {\bibinfo {author} {\bibfnamefont {H.-Y.}\ \bibnamefont
  {Chen}}\ and\ \bibinfo {author} {\bibfnamefont {D.~E.}\ \bibnamefont
  {Holz}},\ }\href@noop {} {\  (\bibinfo {year} {2016})},\ \Eprint
  {http://arxiv.org/abs/1612.01471} {arXiv:1612.01471 [astro-ph.HE]}
  \BibitemShut {NoStop}%
%%CITATION = ARXIV:1612.01471;%%
\bibitem [{\citenamefont {Bertacca}\ \emph {et~al.}(2017)\citenamefont
  {Bertacca}, \citenamefont {Raccanelli}, \citenamefont {Bartolo},\ and\
  \citenamefont {Matarrese}}]{Bertacca:2017vod}%
  \BibitemOpen
  \bibfield  {author} {\bibinfo {author} {\bibfnamefont {D.}~\bibnamefont
  {Bertacca}}, \bibinfo {author} {\bibfnamefont {A.}~\bibnamefont
  {Raccanelli}}, \bibinfo {author} {\bibfnamefont {N.}~\bibnamefont {Bartolo}},
  \ and\ \bibinfo {author} {\bibfnamefont {S.}~\bibnamefont {Matarrese}},\
  }\href@noop {} {\  (\bibinfo {year} {2017})},\ \Eprint
  {http://arxiv.org/abs/1702.01750} {arXiv:1702.01750 [gr-qc]} \BibitemShut
  {NoStop}%
%%CITATION = ARXIV:1702.01750;%%
\bibitem [{\citenamefont {{Gordon}}\ \emph {et~al.}(2007)\citenamefont
  {{Gordon}}, \citenamefont {{Land}},\ and\ \citenamefont
  {{Slosar}}}]{Gordon2007PRL}%
  \BibitemOpen
  \bibfield  {author} {\bibinfo {author} {\bibfnamefont {C.}~\bibnamefont
  {{Gordon}}}, \bibinfo {author} {\bibfnamefont {K.}~\bibnamefont {{Land}}}, \
  and\ \bibinfo {author} {\bibfnamefont {A.}~\bibnamefont {{Slosar}}},\ }\href
  {\doibase 10.1103/PhysRevLett.99.081301} {\bibfield  {journal} {\bibinfo
  {journal} {Physical Review Letters}\ }\textbf {\bibinfo {volume} {99}},\
  \bibinfo {eid} {081301} (\bibinfo {year} {2007})},\ \Eprint
  {http://arxiv.org/abs/0705.1718} {arXiv:0705.1718} \BibitemShut {NoStop}%
\bibitem [{\citenamefont {Nishizawa}\ \emph {et~al.}(2011)\citenamefont
  {Nishizawa}, \citenamefont {Taruya},\ and\ \citenamefont
  {Saito}}]{Nishizawa:2010xx}%
  \BibitemOpen
  \bibfield  {author} {\bibinfo {author} {\bibfnamefont {A.}~\bibnamefont
  {Nishizawa}}, \bibinfo {author} {\bibfnamefont {A.}~\bibnamefont {Taruya}}, \
  and\ \bibinfo {author} {\bibfnamefont {S.}~\bibnamefont {Saito}},\ }\href
  {\doibase 10.1103/PhysRevD.83.084045} {\bibfield  {journal} {\bibinfo
  {journal} {Phys. Rev.}\ }\textbf {\bibinfo {volume} {D83}},\ \bibinfo {pages}
  {084045} (\bibinfo {year} {2011})},\ \Eprint {http://arxiv.org/abs/1011.5000}
  {arXiv:1011.5000 [astro-ph.CO]} \BibitemShut {NoStop}%
\bibitem [{\citenamefont {Tuyenbayev}\ \emph {et~al.}(2017)\citenamefont
  {Tuyenbayev} \emph {et~al.}}]{Tuyenbayev:2016CQG}%
  \BibitemOpen
  \bibfield  {author} {\bibinfo {author} {\bibfnamefont {D.}~\bibnamefont
  {Tuyenbayev}} \emph {et~al.},\ }\href {\doibase
  10.1088/0264-9381/34/1/015002} {\bibfield  {journal} {\bibinfo  {journal}
  {Class. Quant. Grav.}\ }\textbf {\bibinfo {volume} {34}},\ \bibinfo {pages}
  {015002} (\bibinfo {year} {2017})},\ \Eprint
  {http://arxiv.org/abs/1608.05134} {arXiv:1608.05134 [astro-ph.IM]}
  \BibitemShut {NoStop}%
%%CITATION = ARXIV:1608.05134;%%
\bibitem [{\citenamefont {Singer}\ \emph {et~al.}(2016)\citenamefont {Singer}
  \emph {et~al.}}]{Singer2016ApJL}%
  \BibitemOpen
  \bibfield  {author} {\bibinfo {author} {\bibfnamefont {L.~P.}\ \bibnamefont
  {Singer}} \emph {et~al.},\ }\href {\doibase 10.3847/2041-8205/829/1/L15}
  {\bibfield  {journal} {\bibinfo  {journal} {Astrophys. J.}\ }\textbf
  {\bibinfo {volume} {829}},\ \bibinfo {pages} {L15} (\bibinfo {year}
  {2016})},\ \Eprint {http://arxiv.org/abs/1603.07333} {arXiv:1603.07333
  [astro-ph.HE]} \BibitemShut {NoStop}%
%%CITATION = ARXIV:1603.07333;%%
\end{thebibliography}%

\end{document}